\pdfoutput=1
\documentclass[aps,prl,reprint,showpacs,superscriptaddress,floatfix,preprintnumbers]{revtex4-1}

\usepackage{graphicx}
\usepackage{dcolumn}
\usepackage{subfigure}
\usepackage{placeins}
\usepackage[colorlinks,breaklinks]{hyperref}

\newcommand{\ppbar}{$p\bar{p}$}
\newcommand{\roots}{$\sqrt{s}$}
\newcommand{\fb}{fb$^{-1}$}
\newcommand{\gev}{GeV}
\newcommand{\gevc}{GeV/$c$}
\newcommand{\gevcc}{GeV/$c^2$}
\newcommand{\pt}{$p_T$}
\newcommand{\et}{$E_T$}
\newcommand{\ptgg}{$p_T^{\gamma\gamma}$}
\newcommand{\hf}{$h_f$}
\newcommand{\Mh}{$m_H$}
\newcommand{\Mhf}{$m_{h_f}$}
\newcommand{\Hgg}{$H \rightarrow \gamma \gamma$}

\newcommand{\BRHgg}{$\mathcal{B}$($H  \rightarrow \gamma \gamma$)}
\newcommand{\BRHfgg}{$\mathcal{B}$($h_f \rightarrow \gamma \gamma$)}
\newcommand{\py}{\textsc{pythia}}
\newcommand{\mc}{\multicolumn}

\begin{document}

\preprint{FERMILAB-PUB-11-482-E-PPD}

\title{Search for a Higgs Boson in the Diphoton Final State \\ in \ppbar\ Collisions at \roots~=~1.96~TeV}

\affiliation{Institute of Physics, Academia Sinica, Taipei, Taiwan 11529, Republic of China} 
\affiliation{Argonne National Laboratory, Argonne, Illinois 60439, USA} 
\affiliation{University of Athens, 157 71 Athens, Greece} 
\affiliation{Institut de Fisica d'Altes Energies, ICREA, Universitat Autonoma de Barcelona, E-08193, Bellaterra (Barcelona), Spain} 
\affiliation{Baylor University, Waco, Texas 76798, USA} 
\affiliation{Istituto Nazionale di Fisica Nucleare Bologna, $^{aa}$University of Bologna, I-40127 Bologna, Italy} 
\affiliation{University of California, Davis, Davis, California 95616, USA} 
\affiliation{University of California, Los Angeles, Los Angeles, California 90024, USA} 
\affiliation{Instituto de Fisica de Cantabria, CSIC-University of Cantabria, 39005 Santander, Spain} 
\affiliation{Carnegie Mellon University, Pittsburgh, Pennsylvania 15213, USA} 
\affiliation{Enrico Fermi Institute, University of Chicago, Chicago, Illinois 60637, USA}
\affiliation{Comenius University, 842 48 Bratislava, Slovakia; Institute of Experimental Physics, 040 01 Kosice, Slovakia} 
\affiliation{Joint Institute for Nuclear Research, RU-141980 Dubna, Russia} 
\affiliation{Duke University, Durham, North Carolina 27708, USA} 
\affiliation{Fermi National Accelerator Laboratory, Batavia, Illinois 60510, USA} 
\affiliation{University of Florida, Gainesville, Florida 32611, USA} 
\affiliation{Laboratori Nazionali di Frascati, Istituto Nazionale di Fisica Nucleare, I-00044 Frascati, Italy} 
\affiliation{University of Geneva, CH-1211 Geneva 4, Switzerland} 
\affiliation{Glasgow University, Glasgow G12 8QQ, United Kingdom} 
\affiliation{Harvard University, Cambridge, Massachusetts 02138, USA} 
\affiliation{Division of High Energy Physics, Department of Physics, University of Helsinki and Helsinki Institute of Physics, FIN-00014, Helsinki, Finland} 
\affiliation{University of Illinois, Urbana, Illinois 61801, USA} 
\affiliation{The Johns Hopkins University, Baltimore, Maryland 21218, USA} 
\affiliation{Institut f\"{u}r Experimentelle Kernphysik, Karlsruhe Institute of Technology, D-76131 Karlsruhe, Germany} 
\affiliation{Center for High Energy Physics: Kyungpook National University, Daegu 702-701, Korea; Seoul National
University, Seoul 151-742, Korea; Sungkyunkwan University, Suwon 440-746, Korea; Korea Institute of Science and
Technology Information, Daejeon 305-806, Korea; Chonnam National University, Gwangju 500-757, Korea; Chonbuk
National University, Jeonju 561-756, Korea}
\affiliation{Ernest Orlando Lawrence Berkeley National Laboratory, Berkeley, California 94720, USA} 
\affiliation{University of Liverpool, Liverpool L69 7ZE, United Kingdom} 
\affiliation{University College London, London WC1E 6BT, United Kingdom} 
\affiliation{Centro de Investigaciones Energeticas Medioambientales y Tecnologicas, E-28040 Madrid, Spain} 
\affiliation{Massachusetts Institute of Technology, Cambridge, Massachusetts 02139, USA} 
\affiliation{Institute of Particle Physics: McGill University, Montr\'{e}al, Qu\'{e}bec, Canada H3A~2T8;
Simon Fraser University, Burnaby, British Columbia, Canada V5A~1S6; University of Toronto, Toronto, Ontario, Canada M5S~1A7;
and TRIUMF, Vancouver, British Columbia, Canada V6T~2A3} 
\affiliation{University of Michigan, Ann Arbor, Michigan 48109, USA} 
\affiliation{Michigan State University, East Lansing, Michigan 48824, USA}
\affiliation{Institution for Theoretical and Experimental Physics, ITEP, Moscow 117259, Russia}
\affiliation{University of New Mexico, Albuquerque, New Mexico 87131, USA} 
\affiliation{Northwestern University, Evanston, Illinois 60208, USA} 
\affiliation{The Ohio State University, Columbus, Ohio 43210, USA} 
\affiliation{Okayama University, Okayama 700-8530, Japan} 
\affiliation{Osaka City University, Osaka 588, Japan} 
\affiliation{University of Oxford, Oxford OX1 3RH, United Kingdom} 
\affiliation{Istituto Nazionale di Fisica Nucleare, Sezione di Padova-Trento, $^{bb}$University of Padova, I-35131 Padova, Italy} 
\affiliation{LPNHE, Universite Pierre et Marie Curie/IN2P3-CNRS, UMR7585, Paris, F-75252 France} 
\affiliation{University of Pennsylvania, Philadelphia, Pennsylvania 19104, USA}
\affiliation{Istituto Nazionale di Fisica Nucleare Pisa, $^{cc}$University of Pisa, $^{dd}$University of Siena and $^{ee}$Scuola Normale Superiore, I-56127 Pisa, Italy} 
\affiliation{University of Pittsburgh, Pittsburgh, Pennsylvania 15260, USA} 
\affiliation{Purdue University, West Lafayette, Indiana 47907, USA} 
\affiliation{University of Rochester, Rochester, New York 14627, USA} 
\affiliation{The Rockefeller University, New York, New York 10065, USA} 
\affiliation{Istituto Nazionale di Fisica Nucleare, Sezione di Roma 1, $^{ff}$Sapienza Universit\`{a} di Roma, I-00185 Roma, Italy} 
\affiliation{Rutgers University, Piscataway, New Jersey 08855, USA} 
\affiliation{Texas A\&M University, College Station, Texas 77843, USA} 
\affiliation{Istituto Nazionale di Fisica Nucleare Trieste/Udine, I-34100 Trieste, $^{gg}$University of Udine, I-33100 Udine, Italy} 
\affiliation{University of Tsukuba, Tsukuba, Ibaraki 305, Japan} 
\affiliation{Tufts University, Medford, Massachusetts 02155, USA} 
\affiliation{University of Virginia, Charlottesville, Virginia 22906, USA}
\affiliation{Waseda University, Tokyo 169, Japan} 
\affiliation{Wayne State University, Detroit, Michigan 48201, USA} 
\affiliation{University of Wisconsin, Madison, Wisconsin 53706, USA} 
\affiliation{Yale University, New Haven, Connecticut 06520, USA} 
\author{T.~Aaltonen}
\affiliation{Division of High Energy Physics, Department of Physics, University of Helsinki and Helsinki Institute of Physics, FIN-00014, Helsinki, Finland}
\author{B.~\'{A}lvarez~Gonz\'{a}lez$^w$}
\affiliation{Instituto de Fisica de Cantabria, CSIC-University of Cantabria, 39005 Santander, Spain}
\author{S.~Amerio}
\affiliation{Istituto Nazionale di Fisica Nucleare, Sezione di Padova-Trento, $^{bb}$University of Padova, I-35131 Padova, Italy} 
\author{D.~Amidei}
\affiliation{University of Michigan, Ann Arbor, Michigan 48109, USA}
\author{A.~Anastassov}
\affiliation{Northwestern University, Evanston, Illinois 60208, USA}
\author{A.~Annovi}
\affiliation{Laboratori Nazionali di Frascati, Istituto Nazionale di Fisica Nucleare, I-00044 Frascati, Italy}
\author{J.~Antos}
\affiliation{Comenius University, 842 48 Bratislava, Slovakia; Institute of Experimental Physics, 040 01 Kosice, Slovakia}
\author{G.~Apollinari}
\affiliation{Fermi National Accelerator Laboratory, Batavia, Illinois 60510, USA}
\author{J.A.~Appel}
\affiliation{Fermi National Accelerator Laboratory, Batavia, Illinois 60510, USA}
\author{A.~Apresyan}
\affiliation{Purdue University, West Lafayette, Indiana 47907, USA}
% ...
\collaboration{CDF Collaboration}
\thanks{With visitors from $^a$Istituto Nazionale di Fisica Nucleare, Sezione di Cagliari, 09042 Monserrato (Cagliari), Italy,
$^b$University of California Irvine, Irvine, CA  92697, USA,
$^c$University of California Santa Barbara, Santa Barbara, CA 93106, USA,
$^d$University of California Santa Cruz, Santa Cruz, CA  95064, USA,
$^e$CERN,CH-1211 Geneva, Switzerland,
$^f$Cornell University, Ithaca, NY  14853, USA, 
$^g$University of Cyprus, Nicosia CY-1678, Cyprus, 
$^h$Office of Science, U.S. Department of Energy, Washington, DC 20585, USA,
$^i$University College Dublin, Dublin 4, Ireland,
$^j$University of Fukui, Fukui City, Fukui Prefecture, Japan 910-0017,
$^k$Universidad Iberoamericana, Mexico D.F., Mexico,
$^l$Iowa State University, Ames, IA  50011, USA,
$^m$University of Iowa, Iowa City, IA  52242, USA,
$^n$Kinki University, Higashi-Osaka City, Japan 577-8502,
$^o$Kansas State University, Manhattan, KS 66506, USA,
$^p$University of Manchester, Manchester M13 9PL, United Kingdom,
$^q$Queen Mary, University of London, London, E1 4NS, United Kingdom,
$^r$University of Melbourne, Victoria 3010, Australia,
$^s$Muons, Inc., Batavia, IL 60510, USA,
$^t$Nagasaki Institute of Applied Science, Nagasaki, Japan, 
$^u$National Research Nuclear University, Moscow, Russia,
$^v$University of Notre Dame, Notre Dame, IN 46556, USA,
$^w$Universidad de Oviedo, E-33007 Oviedo, Spain, 
$^x$Texas Tech University, Lubbock, TX  79609, USA,
$^y$Universidad Tecnica Federico Santa Maria, 110v Valparaiso, Chile,
$^z$Yarmouk University, Irbid 211-63, Jordan,
$^{hh}$On leave from J.~Stefan Institute, Ljubljana, Slovenia.
}
\noaffiliation

\date{September 21, 2011}

\begin{abstract}
A search for a narrow Higgs boson resonance in the diphoton mass spectrum is 
presented based on data corresponding to 7.0~\fb\ of integrated luminosity from 
\ppbar\ collisions at \roots~= 1.96 TeV collected by the CDF experiment. No 
evidence of such a resonance is observed, and upper limits are set on the cross 
section times branching ratio of the resonant state as a function of Higgs boson 
mass. The limits are interpreted in the context of the standard model and one 
fermiophobic benchmark model where the data exclude fermiophobic Higgs 
bosons with masses below 114~\gevcc\ at a 95\% Bayesian credibility level.
\end{abstract}

% 12.38.Qk	Experimental tests
% 13.85.Rm	Limits on production of particles
% 14.80.Bn	Standard-model Higgs bosons
% 14.80.Ec	Other neutral Higgs bosons
% 12.60.Fr	Extensions of electroweak Higgs sector 
\pacs{12.38.Qk, 13.85.Rm, 14.80.Bn, 14.80.Ec, 12.60.Fr}

\maketitle

%================================================

In the standard model (SM) of particle physics, the electromagnetic and weak 
forces are unified into a single theory known as electroweak theory. However, the 
measured cross sections for electromagnetic and weak interactions differ by 
several orders of magnitude due to massive $W$ and $Z$ bosons that mediate 
the weak interactions. These bosons gain mass via electroweak symmetry 
breaking by way of the Higgs mechanism~\cite{Higgs:1964pj, *Guralnik:1964eu, 
*Englert:1964et}, and the electroweak theory predicts the existence of a boson, 
known as the Higgs boson, that provides a direct test of the theory.

The SM prediction for the Higgs boson branching ratio into a photon pair 
\BRHgg\ is extremely small, reaching a maximal value of only about 0.2\% 
for a Higgs boson mass \Mh~= 120~\gevcc~\cite{Djouadi:1997yw}. Even so, 
a search using the diphoton final state is appealing due to its better mass 
resolution and reconstruction efficiency relative to dominant decay modes 
involving $b$ quarks. The \Hgg\ channel provides its greatest sensitivity for 
Higgs boson masses between 110 and 140~\gevcc, contributing in a region 
most useful to combined Tevatron Higgs boson searches~\cite{Tev:2011cb} 
and overlapping with a region preferred by electroweak 
constraints~\cite{EwkLimits2010}. In addition, in ``fermiophobic'' Higgs
boson models, where the coupling of the Higgs boson to fermions is
suppressed, the diphoton decay can be greatly enhanced~\cite{Rosca:2002me,
*Akeroyd:1995hg, *Portugal, *Mrenna, *Haber:1978jt, *Gunion:1989ci, *Barger:1992ty,
*Basdevant:1992nb, *PhysRevD.49.1354, *Diaz:1994pk, *bruchersantos,
*Matchev1}.

The Collider Detector at Fermilab (CDF) and D0 experiments at the Tevatron 
have searched for both a SM Higgs boson and a fermiophobic Higgs boson 
\hf\ decaying to two photons~\cite{Affolder:2001hx, Abbott:1998vv, hf_D0}. 
The D0 experiment recently set 95\% confidence level (C.L.) upper limits on 
the cross section times branching ratio $\sigma$~$\times$ \BRHgg\ relative 
to the SM prediction and on \BRHfgg\ using data corresponding to an 
integrated luminosity $\cal{L}$ of 8.2~\fb~\cite{d0Higgs2011}. The \hf\ result 
sets a lower limit on \Mhf\ of 112.9~\gevcc, a more stringent limit than that of 
109.7~\gevcc\ obtained from combined searches at the LEP
electron-positron collider at CERN~\cite{Rosca:2002me}. Previously, the
CDF experiment set 95\% C.L. upper limits on \BRHfgg\ with data
corresponding to $\cal{L}$~=~3.0~\fb, resulting in an exclusion of \Mhf\ below
106~\gevcc~\cite{Aaltonen:2009ga}.

In this Letter, we present a search of the diphoton mass distribution from 
CDF data for a narrow resonance that could reveal the presence of a SM or 
fermiophobic Higgs boson. This analysis, which uses more than twice the 
integrated luminosity of the previous CDF \hf\ analysis~\cite{Aaltonen:2009ga},
implements new techniques to improve the identification of 
photons and yields a new, improved lower limit on the fermiophobic Higgs 
boson mass. In addition, this is the first search for the SM Higgs boson at 
CDF using \Hgg\ decays from Run II data.

The SM Higgs production mechanisms considered in this study are gluon 
fusion (GF), associated production (VH) where a Higgs boson is produced in 
association with a $W$ or $Z$ boson, and vector boson fusion (VBF) with 
cross sections of 1072.3~fb~\cite{Anastasiou:2008tj, *deFlorian:2009hc}, 
240.3~fb~\cite{Baglio:2010um, *Brein:2003wg, *Ciccolini:2003jy}, and 
72.7~fb~\cite{Bolzoni:2010xr, *Ciccolini:2007jr, *Ciccolini:2007ec}, 
respectively, for \Mh~= 120~\gevcc. A benchmark fermiophobic model is 
considered in which the Higgs boson does not couple to fermions, yet 
retains its SM couplings to bosons~\cite{Rosca:2002me, *Akeroyd:1995hg,
*Portugal, *Mrenna, *Haber:1978jt, *Gunion:1989ci, *Barger:1992ty,
*Basdevant:1992nb, *PhysRevD.49.1354, *Diaz:1994pk, *bruchersantos,
*Matchev1}. In this model, the GF process is suppressed and the
fermiophobic Higgs boson production is dominated by VH and VBF. 
Furthermore, Higgs boson decays to fermions are removed, resulting in
increased branching ratios for decays into gauge bosons.

We use the CDF II detector~\cite{Acosta:2004yw} to identify photon 
candidate events produced in \ppbar\ collisions at \roots~= 1.96 TeV. The 
silicon vertex tracker~\cite{Sill:2000zz} and the central outer
tracker~\cite{Affolder:2003ep}, contained within a 1.4~T axial magnetic field, 
measure the trajectories of charged particles and determine their momenta. 
Particles that pass through the central outer tracker reach the electromagnetic (EM) and 
hadronic calorimeters~\cite{ccal, wha, Albrow:2001jw}, which are divided 
into two regions: central ($|\eta|<1.1$)~\footnote{CDF uses a cylindrical 
coordinate system with $+z$ in the proton beam direction. $\theta$ and
$\phi$ are the polar and azimuthal angles, respectively, and pseudorapidity
is $\eta = - \ln \tan(\theta/2)$.} and forward or ``plug'' ($1.1<|\eta|<3.6$). The
EM calorimeters contain fine-grained shower maximum
detectors~\cite{Apollinari:1998bg}, which measure the shower shape
and centroid position in the two dimensions transverse to the direction of
the shower development.
 
Events with two photon candidates are selected, and the data are divided 
into four categories according to the position and type of the 
photons. In central-central (CC) events, both photon candidates are located 
within the fiducial region of the central EM calorimeter ($|\eta|< 1.05$); in 
central-plug (CP) events, one photon candidate is located in this region and 
the other is in the fiducial region of the plug calorimeter ($1.2<|\eta|<2.8$); in 
central-central events with a conversion (C$^\prime$C), both photon 
candidates are in the central region, but one photon converts and is 
reconstructed from its $e^+e^-$ decay products; and, in central-plug events with 
a conversion (C$^\prime$P), there is one central conversion candidate 
together with a plug photon candidate.

The events are selected by a three-level trigger system that requires an 
isolated cluster of energy deposited in the EM calorimeter with a transverse 
energy \et~$>25$~\gev~\footnote{The transverse energy \et\ and transverse 
momentum \pt\ are defined as $E \sin\theta$ and $|{\vec{p}}| \sin\theta$, 
respectively.}. The trigger efficiency for events accepted into the final sample 
is determined from simulation and found to be essentially 100\% for the most 
sensitive event category (CC) and above 90\% for all other categories. 

A set of selection criteria is used to remove background events and to 
identify high-energy photon candidates for this analysis. All reconstructed 
photon candidates are required to have \et~$>15$~\gev. Plug photon 
candidates are identified using standard CDF requirements described 
elsewhere~\cite{RS_PRL,Wynne:2009}. A new neural network (NN) technique is 
used to identify photons in the central region. Central photon
candidates are first required to satisfy loose 
selection requirements, as described in Ref.~\cite{Aaltonen:2009in}. 
After additional track requirements are applied to remove electrons, the 
remaining candidates are required to have a NN output value above a 
threshold that is selected to maximize \Hgg\ sensitivity. As more than half of 
the events in the data with two photon candidates contain either one or two 
jets misidentified as a prompt photon~\footnote{Typically, this occurs when a 
jet fragments into a $\pi^0$ or $\eta$ particle that subsequently decays to 
multiple photons, which are then reconstructed as a single photon.}, the NN 
discriminant is trained using photon and jet Monte Carlo (MC) samples and 
constructed from several detector variables that distinguish true photons 
from these jet backgrounds~\footnote{The variables also allow the NN 
method to be applied to electrons, which are used to calibrate ID 
efficiencies.}. These variables include the ratio of energy in the shower 
maximum detector to that in the calorimeter cluster associated with the 
photon, the ratio of hadronic to EM transverse energy, calorimeter 
and track isolation~\cite{Aaltonen:2009in}, and a $\chi^2$ value calculated 
by comparing the measured transverse shower profile to that of a single EM 
shower~\cite{Abe:1993qb}. This NN method increases the photon signal 
efficiency by $\sim$5\% and background rejection by $\sim$12\% compared 
to the standard selection requirements for central 
photons~\cite{Aaltonen:2009in}, which improves \Hgg\ sensitivity by 
about 9\%.

As photons pass through the CDF detector material, EM interactions with a 
nucleus cause about 15\% of central photons to convert into an
electron-positron pair. In order to recover these conversion photons,
we search for a central electron with a nearby track corresponding to a
particle of opposite charge. The proximity of the two tracks is first
determined by requiring the transverse distance between the two tracks
to be less than 0.2~cm at the radial location where they are parallel. The
difference in $\cot \theta$ between the two tracks must be less than 0.04,
where $\cot \theta = p_z/p_T$. Backgrounds are further removed
by requiring the ratio of \et\ to \pt\ of the reconstructed conversion photon
to be between 0.1$c$ and 1.9$c$ and calorimeter isolation to be
less than 2.6~GeV, where cut boundaries are 
optimized to maximize \Hgg\ sensitivity. The direction of the conversion 
photon's momentum is obtained by taking the vector sum of the individual 
track momenta. Better \Hgg\ mass resolution is obtained, however, by setting 
the total momentum to be the conversion photon's energy obtained from EM 
calorimeters. Reconstruction of photon conversions in this analysis provides 
an improvement of about 13\% in sensitivity to a Higgs boson
signal~\cite{kbthesis, *Bland:2011dc}.

The above selection criteria define an inclusive diphoton sample for the SM 
Higgs boson search. In order to improve sensitivity for the fermiophobic 
Higgs boson search, the event selection is extended by taking advantage of 
the final-state features present in the VH and VBF processes. Because the 
Higgs boson from these processes will be produced with a $W$ or $Z$ 
boson or with two jets, the transverse momentum of the diphoton system 
\ptgg\ is generally higher relative to the diphoton backgrounds. A 
requirement of \ptgg~$>75$~\gevc\ forms a region of high \hf\ sensitivity, 
retaining roughly 30\% of the signal while removing 99.5\% of the 
background~\cite{Aaltonen:2009ga}. Two lower \ptgg\ regions are 
additionally included and provide about 15\% more \hf\ sensitivity:
\ptgg~$<35$~\gevc\ and $35$~\gevc~$<$ \ptgg~$<75$~\gevc. With
four diphoton categories (CC, CP, C$^\prime$C, and C$^\prime$P) and
three \ptgg\ regions, twelve independent channels are included for the
fermiophobic Higgs boson search. 
 
The efficiency times detector acceptance for signal events given in 
Table~\ref{tab:acc}~\footnote{Additional information, including tables of
values for other Higgs boson masses, is available in Supplemental Material
at [URL].} is calculated using \py~\cite{PYTHIA} MC event
samples, which are generated as described in 
Ref.~\cite{Aaltonen:2009ga}. Corrections in the photon identification
(ID) efficiencies due to imperfections in the detector simulation are derived
using electrons from $Z$ boson decays by comparing
the ID efficiencies obtained from the detector simulation to the ID efficiencies
measured in the data~\cite{Aaltonen:2009ga}. For central
conversions, a study of $Z \rightarrow e^\pm$~+~trident events in data
and MC is used to obtain a systematic uncertainty 
of 7\% on the efficiency of conversion identification, where a trident is 
defined as an electron that radiates a photon via bremsstrahlung which then 
converts to an electron-positron
pair $(e^\mp \gamma \rightarrow e^\mp e^+e^-)$. 

\begin{table}
\caption{Efficiency times detector acceptance ($\epsilon$A) for signal events 
in each event category (CC, CP, C$^\prime$C, and C$^\prime$P) for
\Mh~= 120~\gevcc, as a percentage of the total number of \Hgg\ decays for
each production mechanism. For the \hf\ search, results for VH and VBF 
are shown for the high{\textbar}medium{\textbar}low \ptgg\ regions as described in the text.
\label{tab:acc}}
\begin{ruledtabular}
\begin{tabular}{cdddcc}
& \mc{3}{c}{$H_\mathrm{SM}$ Search} & \mc{2}{c}{$h_f$ Search} \\
$\epsilon$A (\%) &\mc{1}{c}{GF} &\mc{1}{c}{VH} & \mc{1}{c}{VBF} & \mc{1}{c}{VH} & \mc{1}{c}{VBF} \\
\hline
CC 			& 10.0 & 10.2 & 11.0		& 4.8{\textbar}3.8{\textbar}1.9 		& 4.2{\textbar}4.6{\textbar}2.6 \\
CP 			& 12.0 & 10.9 & 11.1		& 4.2{\textbar}4.7{\textbar}2.5 		& 3.6{\textbar}5.0{\textbar}3.0 \\
C$^\prime$C 	& 2.4	   & 2.3 & 2.6		& 1.0{\textbar}0.9{\textbar}0.4 		& 0.9{\textbar}1.0{\textbar}0.6 \\
C$^\prime$P 	& 1.4   &1.2  & 1.2		&0.4{\textbar}0.5{\textbar}0.3 		& 0.3{\textbar}0.5{\textbar}0.3 \\
\hline
Total 		& 25.8 & 24.6 & 25.9 &  \mc{1}{l}{10.4{\textbar}9.9{\textbar}5.1} & 9.0{\textbar}11.1{\textbar}6.5 \\ 
\end{tabular}
\end{ruledtabular}
\end{table}

The largest systematic uncertainties on the expected number of Higgs 
boson events arise from the conversion ID efficiency (7\%), the integrated 
luminosity measurement (6\%), varying the parton distribution functions 
used in \py\ (up to 5\%)~\cite{CTEQ6.1M,Weights}, varying the parameters 
that control the amount of initial- and final-state radiation from the parton 
shower model of \py\ (about $4\%$)~\footnote{We constrain the rate of initial 
state radiation using Drell-Yan events in data.}, and the \py\ modeling of the 
shape of the \ptgg\ distribution for the \hf\ signal (up to 4\%). The latter 
uncertainty is only for VH and VBF used in the fermiophobic search and was 
obtained by studying the effect on the acceptance from the differences in the 
shape of the \ptgg\ distribution from \py\ and from leading-order and
next-to-leading-order calculations~\cite{MrennaQCD}. Finally, we include 
uncertainties from the photon ID efficiency (up to 4\%), the trigger efficiency 
(less than 3\%), and the EM energy scale (less than 1\%). 

\begin{figure}[tb]
\begin{center}
\includegraphics[width=1.0\linewidth]{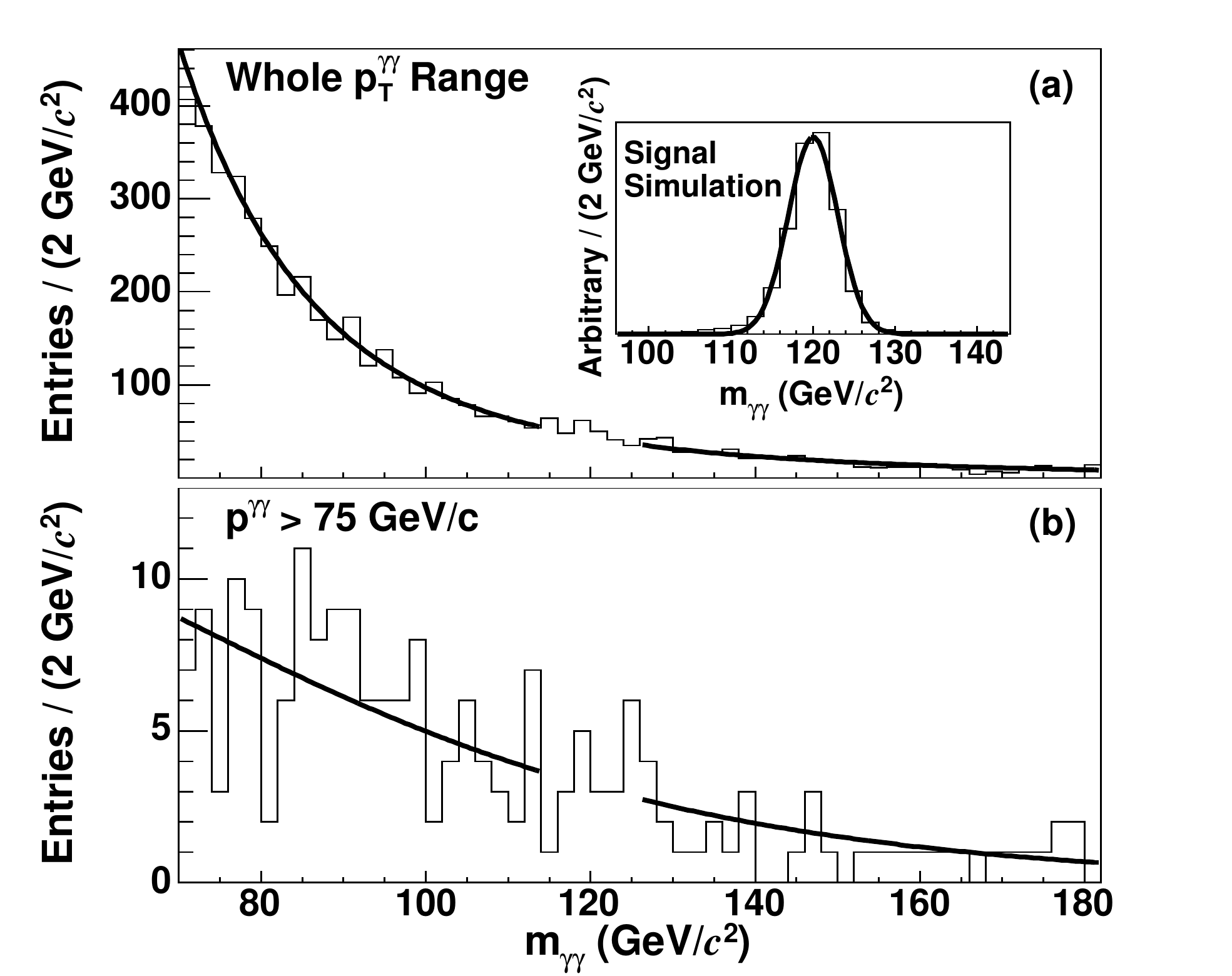}
\end{center}
\caption{The invariant mass distribution of the data for CC photon pairs is 
shown in (a) for the entire \ptgg\ region used in the SM Higgs boson search 
and (b) for the highest \ptgg\ region (the most sensitive region) used in the 
\hf\ search. Each distribution shows a fit to the data for the hypothesis of a 
\Mh\ of 120~\gevcc. The gap in the fit centered at 120~\gevcc\ represents 
the signal region for this mass point that was excluded from the fit. The 
expected shape of the signal from simulation is shown in the inset of (a).
\label{fig:Error_Bands}}
\end{figure}

\begin{table*}
\caption{Expected and observed 95\% C.L. upper limits on the production 
cross section times branching ratio relative to the SM prediction, the 
production cross section times branching ratio with theoretical cross section 
uncertainties removed, and the \hf\ branching ratio. The fermiophobic 
benchmark model prediction for \BRHfgg\ is also shown for comparison. 
\label{tab:limits}}
\begin{ruledtabular}
\begin{tabular}{ccddddddddddd}
& \Mh\ (\gevcc) & \mc{1}{c}{100} & \mc{1}{c}{105} & \mc{1}{c}{110} &
 \mc{1}{c}{115} & \mc{1}{c}{120} & \mc{1}{c}{125} &
 \mc{1}{c}{130} & \mc{1}{c}{135} & \mc{1}{c}{140} &
 \mc{1}{c}{145} & \mc{1}{c}{150} \\
\hline
$\sigma\times$ \BRHgg/SM & Expected & 16.4& 14.8& 14.2& 13.8& 13.3& 13.6& 14.4& 15.8& 17.7& 20.8& 27.5 \\
& Observed & 15.1& 13.9& 8.5& 14.6& 28.7& 19.2& 19.2& 14.8& 23.1& 21.9& 21.4 \\
\hline
$\sigma\times$ \BRHgg\ (fb) &Expected & 57.3& 50.8& 47.9& 43.2& 39.0& 36.0& 32.8& 30.6& 28.5& 26.7& 26.1 \\
& Observed & 52.9& 47.8& 28.8& 44.8& 84.4& 50.4& 44.7& 29.4& 36.9& 28.0& 20.2 \\
\hline
& Expected & 4.4& 4.9& 5.2& 5.8& 6.0& 6.4& 6.8& 7.4& 7.7& 8.1& 8.7 \\
\BRHfgg\ (\%) & Observed & 4.8& 5.4& 2.8& 4.2& 7.3& 5.5& 6.6& 6.6& 5.7& 7.8& 8.1 \\
& Fermiophobic Prediction & 18.2 & 10.6 & 6.2 & 3.8 & 2.8 & 2.2 & 1.9 & 1.2 & 0.6 & 0.3 & 0.2 \\
\end{tabular}
\end{ruledtabular}
\end{table*}

\begin{figure*}
\begin{center}
\includegraphics[width=\linewidth]{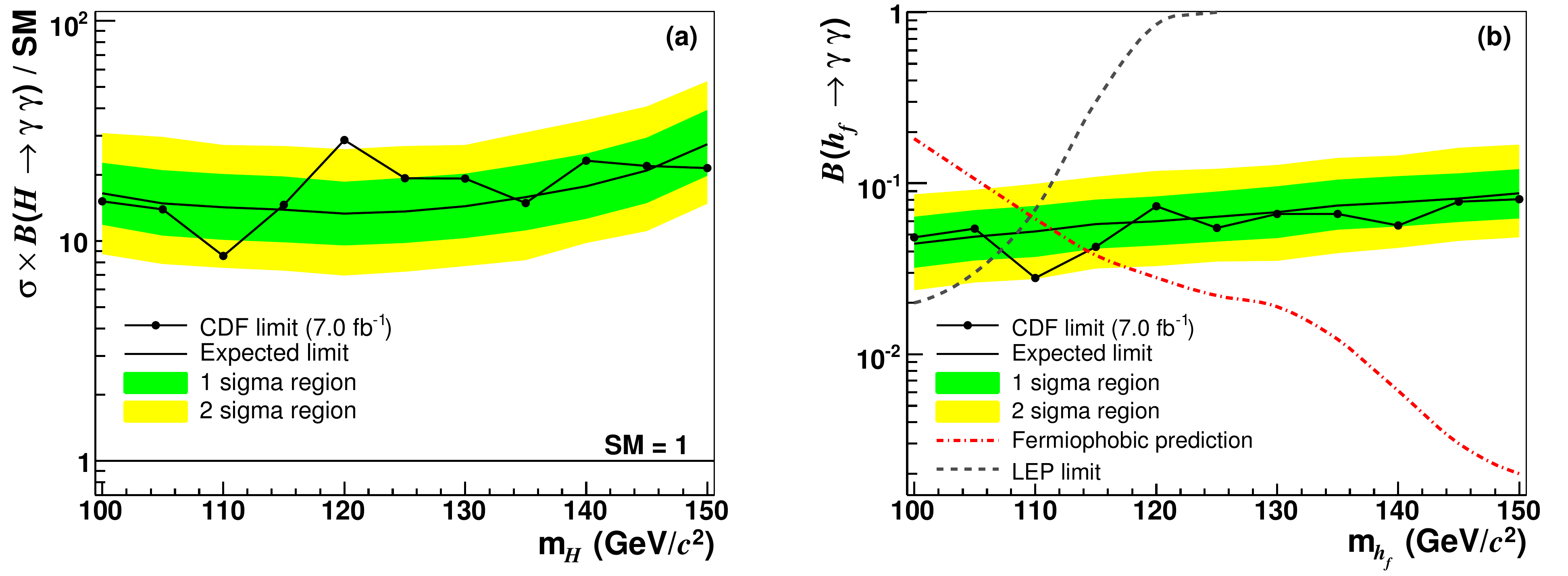}
\end{center}
\caption{
(a) As a function of \Mh, the 95\% C.L. upper limit on cross section times 
branching ratio for the SM Higgs boson decay to two photons, relative to the 
SM prediction. (b) The 95\% C.L. upper limit on the branching ratio for the 
fermiophobic Higgs boson decay to two photons, as a function of \Mhf. For 
reference, the 95\% C.L. limits from LEP are also included. The shaded 
regions represent the 1$\sigma$ and 2$\sigma$ probability of fluctuations of 
the observed limit away from the expected limit based on the distribution of 
simulated experimental outcomes.\label{fig:Limits}} 
\end{figure*}

The decay of a Higgs boson into a photon pair would appear as a very 
narrow peak in the invariant mass distribution of the two photons (see
Fig.~\ref{fig:Error_Bands} as an example for the CC sample). The diphoton
mass resolution, as determined from simulation and checked using
$Z \rightarrow e^+e^-$ decays in data, is better than 3~\gevcc\ for the
Higgs boson mass regions and diphoton channels studied here and is limited
by the energy resolution of the EM calorimeters~\footnote{The natural width
of the Higgs boson is negligible.}. The mass resolution is also sensitive to
the selection of the correct primary vertex of the \ppbar\ interaction, determined
by selecting the vertex with the highest sum of associated track momenta. 
The locations of the vertex and EM energy cluster are used to derive the
photon's momentum. For GF (VH and VBF) signal samples, the primary vertex
is misidentified in roughly 16\% (4\%) of nonconversion channel (CC and CP) 
events, which degrades the resolution of the reconstructed Higgs boson 
mass~\footnote{In CC (CP) events, the resolution degrades by 15 (12)\%.}. 
This effect is studied using $Z$ decays in the data and found to be 
well-modeled in the simulation. 

The total background prediction is estimated from a fit made to the data 
using a binned log-likelihood ($\log \ell$) method~\footnote{The combined 
\ptgg\ data and low-\ptgg\ data are fit to a sum of two exponentials multiplied 
by a fractional-degree polynomial, where the degree of one term is a 
parameter of the fit and the two higher \ptgg\ regions are fit to a simpler third-degree
polynomial times an exponential. Channels with a plug photon have a
non-negligible contamination from $Z$ boson decays and additionally include a 
Breit-Wigner function to model this background.}. The fit is performed for 
each \Mh\ hypothesis in 5~\gevcc\ steps from 100 to 150~\gevcc. At each 
step, a 12~\gevcc\ mass window centered on the point is excluded, where 12 
\gevcc\ is chosen to include 95\% of the signal. Fits for a \Mh\ hypothesis of 
120~\gevcc\ are shown in Fig.~\ref{fig:Error_Bands}. The statistical 
uncertainties on the total background in the signal region, taken from the fit, 
are 8\% or less for the channels associated with the SM Higgs boson search 
and 12\% or less for the channels associated with the fermiophobic Higgs 
boson search (except for the high-\ptgg\ bins with conversion photons, 
where it is 27\%). 

No obvious evidence of a narrow peak or any other anomalous structure is
visible in the diphoton mass spectrum. We calculate a Bayesian C.L. limit
for each Higgs boson mass hypothesis based on a combination of
binned likelihoods for all channels using six bins in the 12~\gevcc\ 
signal region (2~\gevcc\ bin width) of each mass distribution.
We use a flat prior in $\sigma$~$\times$ \BRHgg\ and 
integrate over the priors for the systematic uncertainties. A 95\% C.L. limit is 
determined such that 95\% of the posterior density for $\sigma$~$\times$ 
\BRHgg\ falls below the limit~\cite{Nakamura:2010zz}. The expected 95\% 
C.L. limits are calculated assuming no signal, based on expected 
backgrounds only, as the median of 2000 simulated experiments. The 
observed 95\% C.L. on $\sigma$~$\times$ \BRHgg\ are calculated from the 
data. The limit results are displayed in Table~\ref{tab:limits} and graphically 
in Fig.~\ref{fig:Limits}. For a SM Higgs boson, the results are shown relative 
to the theory prediction, where theoretical cross section uncertainties of
14\% on the GF process, 6\% on the VH process, and 5\% on the VBF process 
are included in the limit calculation~\cite{Botje:2011sn, Dittmaier:2011ti}. 
Limits are also provided on $\sigma$~$\times$ \BRHgg\ without including 
theoretical cross section uncertainties. The inclusion of systematic 
uncertainties in the SM (fermiophobic) limit calculation degrades the limit on 
$\sigma$~$\times$ \BRHgg\ by 15\% (9\%), where the effect of the 
uncertainty on the background estimate is dominant at 10\% (6\%).

For the SM limit at \Mh~=~120~\gevcc, we observe a deviation of
greater than 2.5$\sigma$ from the expectation~\cite{Barate:2003sz}.
After accounting for the trials factor associated with performing the search at 
11 mass points, the significance of this discrepancy decreases to less than 
2$\sigma$. When the analysis is optimized for the fermiophobic benchmark 
model, no excess is observed. For the \hf\ model, SM cross sections and 
uncertainties are assumed (GF excluded) and used to convert limits on 
$\sigma$~$\times$ \BRHfgg\ into limits on \BRHfgg. Table~\ref{tab:limits} 
gives the predicted \BRHfgg\ for this model as calculated using
\textsc{hdecay}~\cite{Djouadi:1997yw}.  We obtain a lower limit on
\Mhf\ of $114$~\gevcc\ by linear interpolation between the sampled values
of \Mhf\ based on the intersection of the observed limit and the model
prediction.

This Letter presents the results of a search for a narrow resonance in the
diphoton mass spectrum using data taken by the CDF II detector at the
Tevatron. We have improved upon the previous CDF analysis by implementing
a neural network discriminant to improve central photon identification,
recovering central photons that have converted to an $e^+e^-$ pair, and
more than doubling the amount of data analyzed. There is no significant
evidence of a resonance in the data. Limits are placed on the production
cross section times branching ratio for Higgs boson decay into a photon
pair and compared to the predictions of the standard model and a benchmark
fermiophobic model. The latter result excludes fermiophobic
Higgs boson masses below 114~\gevcc\ at the 95\% C.L., which is the
strongest limit to date on this model by a single experiment.

\begin{acknowledgments}
We thank the Fermilab staff and the technical staffs of the participating 
institutions for their vital contributions. This work was supported by the U.S. 
Department of Energy and National Science Foundation; the Italian Istituto 
Nazionale di Fisica Nucleare; the Ministry of Education, Culture, Sports, 
Science, and Technology of Japan; the Natural Sciences and Engineering 
Research Council of Canada; the National Science Council of the Republic 
of China; the Swiss National Science Foundation; the A.P. Sloan 
Foundation; the Bundesministerium f\"ur Bildung und Forschung, Germany; 
the Korean World Class University Program and the National Research 
Foundation of Korea; the Science and Technology Facilities Council and 
the Royal Society, UK; the Institut National de Physique Nucleaire et 
Physique des Particules/CNRS of France; the Russian Foundation for Basic 
Research; the Ministerio de Ciencia e Innovaci\'{o}n and Programa 
Consolider-Ingenio 2010, Spain; the Slovak R\&D Agency; the Academy of 
Finland; and the Australian Research Council (ARC). 
\end{acknowledgments}

\clearpage

\onecolumngrid
\setcounter{figure}{0}
\setcounter{table}{0}

\begin{table*}
\begin{center}
{\Large{\bf  Supplemental Material

\vspace{2mm}
for 

\vspace{2mm}
Search for a Higgs Boson in the Diphoton Final State \\ in \ppbar\ Collisions at \roots~=~1.96~TeV}

\vspace{8mm}
\large{
The CDF Collaboration

\vspace{2mm}
}}
\end{center}
\end{table*}

\begin{figure*}[h]
\makebox[\textwidth]{
\subfigure[]{\includegraphics[scale=0.4]{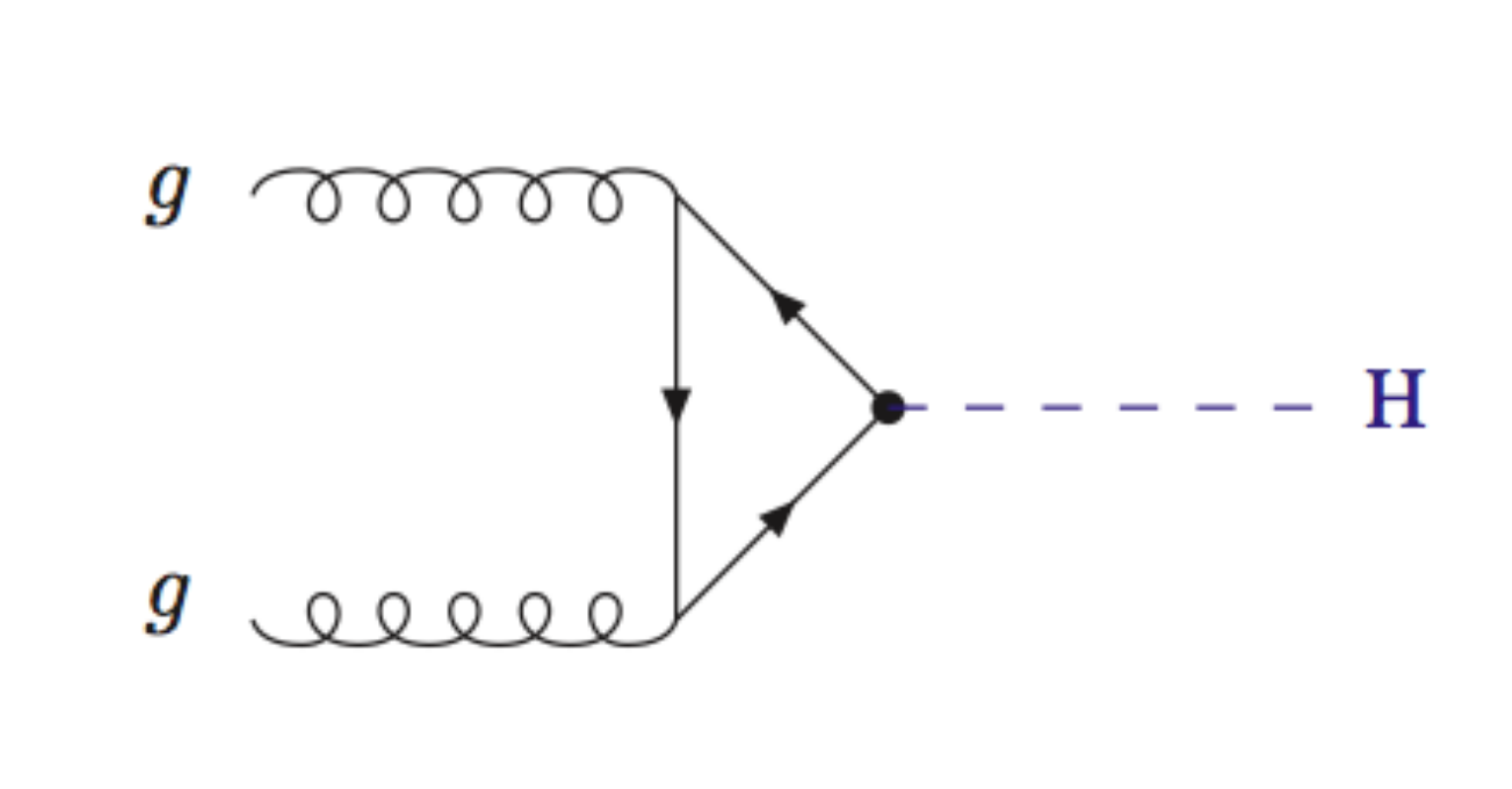}}
\subfigure[]{\includegraphics[scale=0.4]{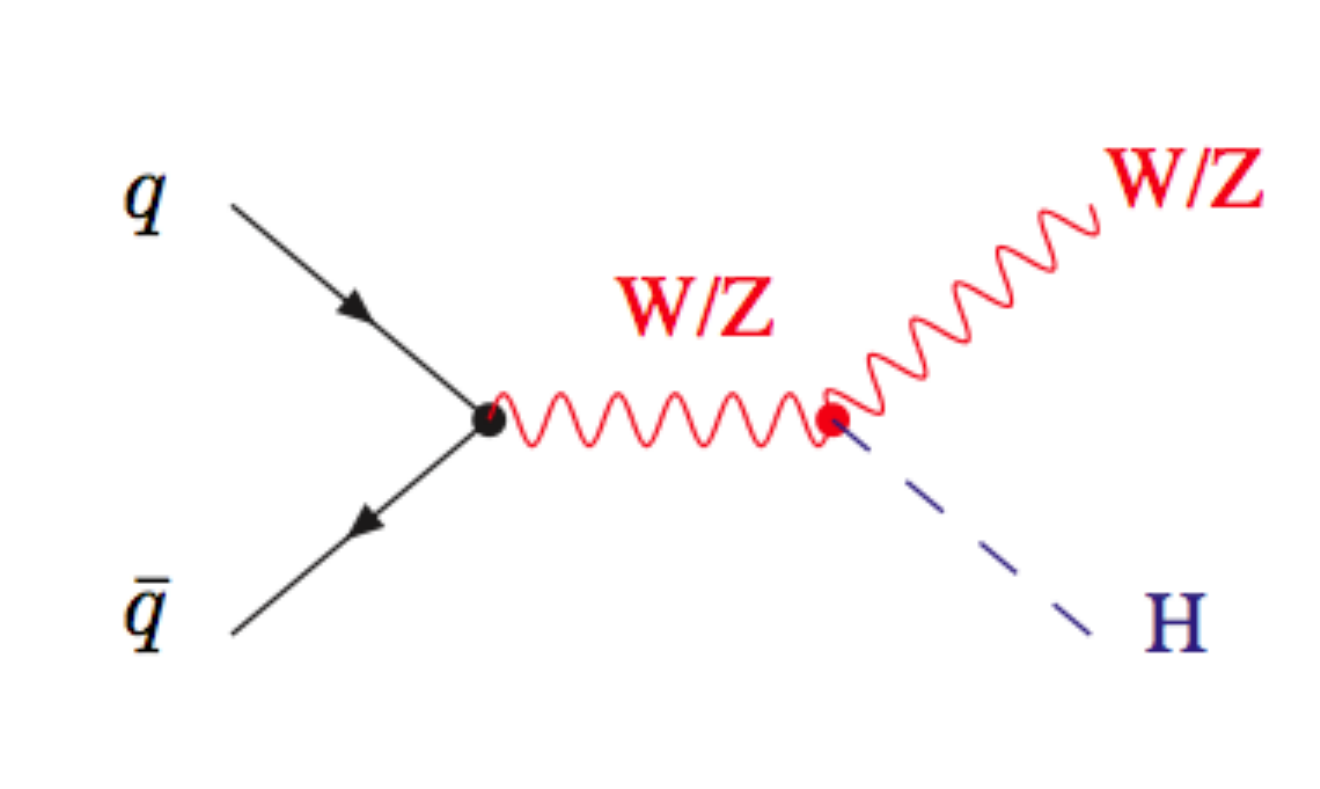}}
\subfigure[]{\includegraphics[scale=0.4]{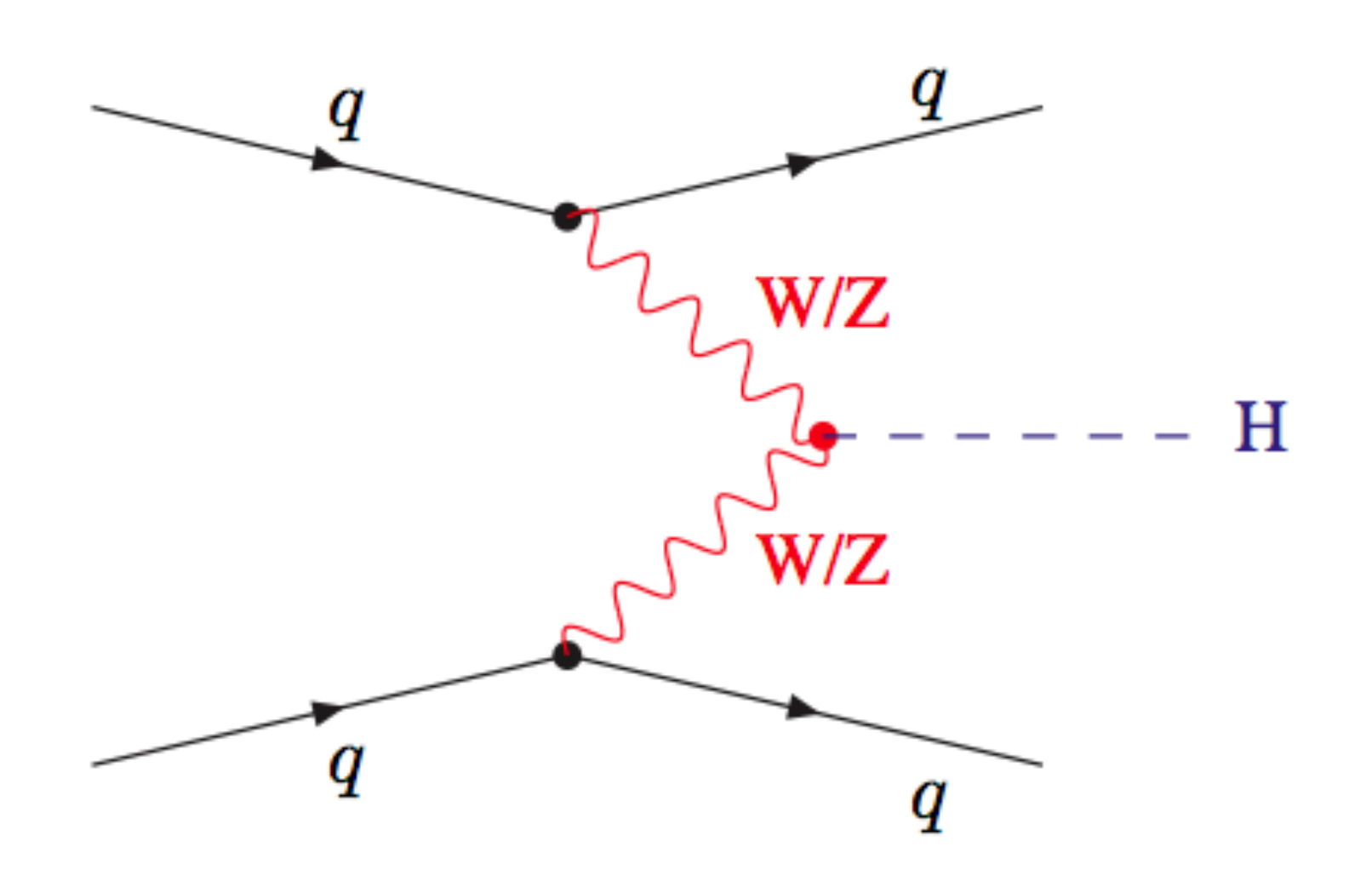}}
}
\caption{The dominant production mechanisms at the Tevatron for the standard model (SM)
 Higgs boson: (a) gluon fusion (GF), (b) associated production (VH) where a Higgs boson is produced in 
association with a $W$ or $Z$ vector boson, and (c) vector boson fusion (VBF).
For the fermiophobic Higgs boson $h_f$ considered in this analysis, SM couplings are assumed;
however, the gluon fusion process is suppressed and is therefore not included.}
\label{production}
\end{figure*}

\begin{table*}[h!]
\caption{Production cross sections for the SM Higgs boson are given for the GF, VH (V~=~$W$ or $Z$), and VBF
production mechanisms shown in Fig.~\ref*{production}. For $h_f$, the GF mechanism is excluded.
The branching ratios for the decay to a photon pair are also 
shown for both the SM and fermiophobic Higgs boson.}
\label{tab:sigma}
\begin{center}
\makebox[\textwidth]{
\begin{tabular}{c@{\quad}c@{\quad}c@{\quad}c@{\quad}c@{\quad}c@{\quad}c}
\hline\hline\\[-2.5ex]
$M_{H}$ (GeV/c$^2$)                        &  
$\sigma_\mathrm{GF}$ (fb)           &
$\sigma_\mathrm{WH}$ (fb)           &
$\sigma_\mathrm{ZH}$ (fb)           & 
$\sigma_\mathrm{VBF}$ (fb)                     & 
$B$($H \rightarrow \gamma \gamma)$ (\%) &
$B$($h_f \rightarrow \gamma \gamma)$ (\%) \\
\hline
100	&	1821.8	&	291.9	&	169.8	&	100.1	&	0.15	&	18.2	\\
105	&	1584.7	&	248.4	&	145.9	&	92.3	&	0.17	&	10.6	\\
110	&	1385.0	&	212.0	&	125.7	&	85.1	&	0.19	&	6.2	\\
115	&	1215.9	&	174.5	&	103.9	&	78.6	&	0.20	&	3.8	\\
120	&	1072.3	&	150.1	&	90.2	&	72.7	&	0.22	&	2.8	\\
125	&	949.3	&	129.5	&	78.5	&	67.1	&	0.22	&	2.2	\\
130	&	842.9	&	112.0	&	68.5	&	62.1	&	0.22	&	1.9	\\
135	&	750.8	&	97.2	&	60.0	&	57.5	&	0.21	&	1.2	\\
140	&	670.6	&	84.6	&	52.7	&	53.2	&	0.19	&	0.6	\\
145	&	600.6	&	73.7	&	46.3	&	49.4	&	0.17	&	0.3	\\
150	&	539.1	&	64.4	&	40.8	&	45.8	&	0.14	&	0.2	\\
\hline                    
\hline
 \end{tabular}}
\end{center}
\end{table*}

%%%%%%%%%%%%%%%%%%%%%%%%%%%%%%%%%%%%%%%%%%%%%%%%%
%                                       SM  Whole Pt Bin
%%%%%%%%%%%%%%%%%%%%%%%%%%%%%%%%%%%%%%%%%%%%%%%%%
\newpage
\section{Standard Model Higgs Boson Search}
        \begin{table}[htb] 
   \caption{The SM Higgs boson search is divided into four independent categories
   (CC, CP, C$^{\prime}$C, and C$^{\prime}$P) as defined in the primary text
   of this Letter. For each SM Higgs boson mass hypotheses tested in this analysis, the efficiency 
   multiplied by signal acceptance ($\epsilon$A) is shown as a percentage of the total number 
   of \Hgg\ decays for each production mechanism (GF, VH, and VBF).
   These values, along with the cross sections and branching ratios provided 
   in Table~\ref*{tab:sigma}, are used to obtain the predicted number of SM Higgs boson
   signal events. The CC and C$^{\prime}$C (CP and C$^{\prime}$P) channels use 
   data corresponding to an integrated luminosity of 7.0~fb$^{-1}$ (6.7~fb$^{-1}$). 
   The number of background and data events are also given for each mass. 
   The final column in each subtable is the number of signal events divided by the
   square root of the number of background events ($S$/$\sqrt{B}$). 
   The event yields for each mass point are 
   obtained from a 12~GeV/$c^{2}$ signal region centered on the Higgs boson mass
   hypothesis, allowing a 2~GeV/$c^{2}$ overlap between signal regions.}
   \label{Tab:EventYields_Whole} 
      \begin{centering}
\makebox[\textwidth]{
      \subtable[]{
     \scalebox{1.0}{ 
      \begin{tabular}{cccccccc} 
      \hline\hline\\[-2ex]
\multicolumn{4}{l}{CC Channel} &   \multicolumn{4}{r}{CDF Run II, 7.0 fb$^{-1}$}    \\
\hline\\[-2.3ex]
      $m_H$ & \multicolumn{3}{c}{$\epsilon$A (\%)} & $H_{SM}$ &  &  & \\
      (GeV/$c^{2}$) & GF & VH & VBF & Signal & Background & Data & $S$/$\sqrt{B}$\\
      \hline
100	&	9.8	&	10.1	&	11.0	&	2.5	&	621	&	615	&	0.10	\\
105	&	9.9	&	10.0	&	11.0	&	2.5	&	526	&	502	&	0.11	\\
110	&	9.9	&	10.2	&	11.0	&	2.4	&	414	&	400	&	0.12	\\
115	&	9.9	&	10.2	&	11.0	&	2.2	&	346	&	361	&	0.12	\\
120	&	10.0	&	10.2	&	11.0	&	2.2	&	271	&	308	&	0.13	\\
125	&	10.0	&	10.2	&	10.9	&	1.9	&	237	&	279	&	0.12	\\
130	&	10.1	&	10.2	&	11.1	&	1.7	&	197	&	207	&	0.12	\\
135	&	10.0	&	10.1	&	10.9	&	1.4	&	177	&	181	&	0.11	\\
140	&	10.2	&	10.3	&	11.1	&	1.2	&	144	&	150	&	0.10	\\
145	&	10.2	&	10.1	&	10.9	&	0.9	&	128	&	129	&	0.08	\\
150	&	10.2	&	10.2	&	11.0	&	0.7	&	112	&	99	&	0.07	\\
      \hline \hline
   \end{tabular} }
   \label{tab:EventYieldsCC} 
  }\hspace{0.6em}
  \subtable[]{
\scalebox{1.0}{ 
      \begin{tabular}{cccccccc} 
      \hline\hline\\[-2ex]
\multicolumn{4}{l}{CP Channel} &   \multicolumn{4}{r}{CDF Run II, 6.7 fb$^{-1}$}    \\
\hline\\[-2.3ex]
      $m_H$ & \multicolumn{3}{c}{$\epsilon$A (\%)} & $H_{SM}$ &  &  & \\
      (GeV/$c^{2}$) & GF & VH & VBF & Signal & Background & Data & $S$/$\sqrt{B}$\\
      \hline
100	&	11.5	&	10.3	&	10.3	&	2.7	&	4296	&	4244	&	0.04	\\
105	&	11.7	&	10.4	&	10.6	&	2.7	&	3580	&	3613	&	0.04	\\
110	&	11.9	&	10.6	&	10.8	&	2.7	&	2919	&	2851	&	0.05	\\
115	&	11.9	&	10.8	&	10.9	&	2.5	&	2547	&	2472	&	0.05	\\
120	&	12.0	&	10.9	&	11.1	&	2.4	&	2071	&	2075	&	0.05	\\
125	&	12.0	&	10.8	&	11.1	&	2.1	&	1819	&	1866	&	0.05	\\
130	&	11.9	&	10.8	&	11.1	&	1.9	&	1511	&	1506	&	0.05	\\
135	&	11.9	&	10.9	&	11.1	&	1.6	&	1332	&	1359	&	0.04	\\
140	&	11.7	&	10.8	&	11.0	&	1.3	&	1118	&	1093	&	0.04	\\
145	&	11.8	&	10.8	&	11.0	&	1.0	&	992	&	964	&	0.03	\\
150	&	11.6	&	10.7	&	10.9	&	0.7	&	825	&	851	&	0.03	\\
      \hline \hline
   \end{tabular}} 
   \label{tab:EventYieldsCP} 
   }}\vspace{5 mm}
   \makebox[\textwidth]{
  \subtable[]{
\scalebox{1.0}{ 
      \begin{tabular}{cccccccc} 
      \hline\hline\\[-2ex]
\multicolumn{4}{l}{C$^{\prime}$C Channel} &   \multicolumn{4}{r}{CDF Run II, 7.0 fb$^{-1}$}    \\
\hline\\[-2.3ex]
      $m_H$ & \multicolumn{3}{c}{$\epsilon$A (\%)} & $H_{SM}$ &  &  & \\
      (GeV/$c^{2}$) & GF & VH & VBF & Signal & Background & Data & $S$/$\sqrt{B}$\\
      \hline
100	&	2.3	&	2.2	&	2.5	&	0.6	&	159	&	158	&	0.05	\\
105	&	2.3	&	2.2	&	2.5	&	0.6	&	126	&	141	&	0.05	\\
110	&	2.3	&	2.3	&	2.6	&	0.6	&	107	&	92	&	0.05	\\
115	&	2.3	&	2.2	&	2.5	&	0.5	&	87.8	&	80	&	0.05	\\
120	&	2.4	&	2.3	&	2.6	&	0.5	&	66.5	&	67	&	0.06	\\
125	&	2.4	&	2.2	&	2.5	&	0.4	&	58.8	&	55	&	0.06	\\
130	&	2.4	&	2.3	&	2.6	&	0.4	&	47.4	&	44	&	0.06	\\
135	&	2.4	&	2.2	&	2.5	&	0.3	&	40.7	&	39	&	0.05	\\
140	&	2.4	&	2.3	&	2.5	&	0.3	&	31.8	&	40	&	0.05	\\
145	&	2.3	&	2.2	&	2.4	&	0.2	&	29.1	&	38	&	0.04	\\
150	&	2.4	&	2.3	&	2.6	&	0.2	&	27.1	&	23	&	0.03	\\
      \hline \hline
   \end{tabular} }
   \label{tab:EventYieldsCCConv} 
   }\hspace{0.6em}
     \subtable[]{
\scalebox{1.0}{ 
      \begin{tabular}{cccccccc} 
      \hline\hline\\[-2ex]
\multicolumn{4}{l}{C$^{\prime}$P Channel} &   \multicolumn{4}{r}{CDF Run II, 6.7 fb$^{-1}$}    \\
\hline\\[-2.3ex]
      $m_H$ & \multicolumn{3}{c}{$\epsilon$A (\%)} & $H_{SM}$ &  &  & \\
      (GeV/$c^{2}$) & GF & VH & VBF & Signal & Background & Data & $S$/$\sqrt{B}$\\
      \hline
100	&	1.3	&	1.0	&	1.1	&	0.3	&	671	&	652	&	0.01	\\
105	&	1.3	&	1.2	&	1.2	&	0.3	&	466	&	443	&	0.01	\\
110	&	1.3	&	1.1	&	1.2	&	0.3	&	328	&	356	&	0.02	\\
115	&	1.3	&	1.1	&	1.2	&	0.3	&	280	&	318	&	0.02	\\
120	&	1.4	&	1.2	&	1.2	&	0.3	&	225	&	268	&	0.02	\\
125	&	1.4	&	1.2	&	1.2	&	0.2	&	204	&	235	&	0.02	\\
130	&	1.4	&	1.2	&	1.3	&	0.2	&	175	&	181	&	0.02	\\
135	&	1.4	&	1.2	&	1.3	&	0.2	&	160	&	140	&	0.01	\\
140	&	1.4	&	1.2	&	1.3	&	0.1	&	138	&	106	&	0.01	\\
145	&	1.4	&	1.2	&	1.3	&	0.1	&	120	&	104	&	0.01	\\
150	&	1.4	&	1.2	&	1.3	&	0.1	&	101	&	99	&	0.01	\\
      \hline \hline
   \end{tabular} }}
   \label{tab:EventYieldsCPConv} 
   }
   \end{centering}
   \end{table} 

%%%%%%%%%%%%%%%%%%%%%%%%%%%%%%%%%%%%%%%%%%%%%%%%%
%                                         High Pt Bin
%%%%%%%%%%%%%%%%%%%%%%%%%%%%%%%%%%%%%%%%%%%%%%%%%
\newpage
\section{Fermiophobic Higgs Boson Search}
\vspace{5mm}
\subsection{(A) $p_T^{\gamma\gamma}>$~75~GeV/c Region}

        \begin{table}[htb] 
   \caption{The fermiophobic Higgs boson search is divided into four independent categories (CC, CP, C$^{\prime}$C, and C$^{\prime}$P) 
   as in the SM Higgs boson search, and it is additionally divided into three regions of 
   $p_T^{\gamma\gamma}$. The $p_T^{\gamma\gamma}>$~75~GeV/c region is shown here,
   which provides the greatest sensitivity for a fermiophobic Higgs boson observation. 
   For each $h_f$ mass hypotheses tested in this analysis, the efficiency 
   multiplied by signal acceptance ($\epsilon$A) is shown as a percentage of the total number 
   of \Hgg\ decays for each production mechanism (VH and VBF).
   These values, along with the cross sections and branching ratios provided 
   in Table~\ref*{tab:sigma}, are used to obtain the predicted number of SM Higgs boson
   signal events. The CC and C$^{\prime}$C (CP and C$^{\prime}$P) channels use 
   data corresponding to an integrated luminosity of 7.0~fb$^{-1}$ (6.7~fb$^{-1}$). 
   The number of background and data events are also given for each mass. 
   The final column in each subtable is the number of signal events divided by the
   square root of the number of background events ($S$/$\sqrt{B}$). 
   The event yields for each mass point are 
   obtained from a 12~GeV/$c^{2}$ signal region centered on the Higgs boson mass
   hypothesis, allowing a 2~GeV/$c^{2}$ overlap between signal regions.}   \label{Tab:EventYields_High} 
      \begin{centering}
\makebox[\textwidth]{
      \subtable[]{
     \scalebox{1.0}{ 
      \begin{tabular}{ccccccc}
      \hline\hline\\[-2ex] 
\multicolumn{4}{l}{CC Channel, $p_T^{\gamma\gamma}>$~75~GeV} & \multicolumn{3}{r}{CDF Run II, 7.0 fb$^{-1}$}    \\
\hline\\[-2.3ex]
      $m_{h_f}$ & \multicolumn{2}{c}{$\epsilon$A (\%)} & $h_{f}$ &  &  & \\
      (GeV/$c^{2}$) & VH & VBF & Signal & Background & Data & $S$/$\sqrt{B}$\\
      \hline
100	&	4.14	&	4.03	&	29.5	&	29.8	&	32	&	5.4	\\
105	&	4.32	&	4.10	&	15.4	&	29.7	&	27	&	2.8	\\
110	&	4.58	&	4.14	&	8.2	&	25.3	&	23	&	1.6	\\
115	&	4.71	&	4.18	&	4.4	&	22.6	&	21	&	0.9	\\
120	&	4.82	&	4.18	&	2.9	&	19.1	&	21	&	0.7	\\
125	&	4.99	&	4.18	&	2.0	&	16.6	&	23	&	0.5	\\
130	&	5.15	&	4.25	&	1.6	&	15.3	&	16	&	0.4	\\
135	&	5.22	&	4.24	&	0.9	&	14.6	&	10	&	0.2	\\
140	&	5.49	&	4.31	&	0.4	&	12.9	&	7	&	0.1	\\
145	&	5.47	&	4.27	&	0.2	&	11.5	&	8	&	0.05	\\
150	&	5.65	&	4.37	&	0.1	&	9.7	&	7	&	0.04	\\ \hline\hline 
   \end{tabular} }
   \label{tab:EventYieldsCC_High} 
  }\hspace{0.6em}
  \subtable[]{
\scalebox{1.0}{ 
      \begin{tabular}{ccccccc} 
      \hline\hline\\[-2ex]
\multicolumn{4}{l}{CP Channel, $p_T^{\gamma\gamma}>$~75~GeV} & \multicolumn{3}{r}{CDF Run II, 6.7 fb$^{-1}$}    \\
\hline\\[-2.3ex]
      $m_{h_f}$ & \multicolumn{2}{c}{$\epsilon$A (\%)} & $h_{f}$ &  &  & \\
      (GeV/$c^{2}$) & VH & VBF & Signal & Background & Data & $S$/$\sqrt{B}$\\
      \hline
100	&	3.18	&	2.97	&	21.5	&	69.3	&	76	&	2.6	\\
105	&	3.36	&	3.09	&	11.4	&	67.6	&	73	&	1.4	\\
110	&	3.60	&	3.32	&	6.2	&	66.0	&	63	&	0.8	\\
115	&	3.96	&	3.37	&	3.5	&	64.0	&	58	&	0.4	\\
120	&	4.20	&	3.58	&	2.4	&	60.6	&	50	&	0.3	\\
125	&	4.40	&	3.62	&	1.7	&	59.0	&	35	&	0.2	\\
130	&	4.55	&	3.72	&	1.3	&	52.4	&	46	&	0.2	\\
135	&	4.78	&	3.79	&	0.8	&	47.9	&	51	&	0.1	\\
140	&	4.98	&	3.80	&	0.4	&	43.3	&	51	&	0.05	\\
145	&	5.15	&	3.89	&	0.2	&	41.0	&	42	&	0.03	\\
150	&	5.21	&	3.89	&	0.1	&	38.1	&	36	&	0.02	\\      \hline\hline
   \end{tabular}} 
   \label{tab:EventYieldsCP_High} 
   }}\vspace{5 mm}
   \makebox[\textwidth]{
  \subtable[]{
\scalebox{1.0}{ 
      \begin{tabular}{ccccccc} 
      \hline\hline\\[-2ex]
\multicolumn{4}{l}{C$^{\prime}$C Channel, $p_T^{\gamma\gamma}>$~75~GeV} &   \multicolumn{3}{r}{CDF Run II, 7.0 fb$^{-1}$}    \\
\hline\\[-2.3ex]
      $m_{h_f}$ & \multicolumn{2}{c}{$\epsilon$A (\%)} & $h_{f}$ &  &  & \\
      (GeV/$c^{2}$) & VH & VBF & Signal & Background & Data & $S$/$\sqrt{B}$\\
      \hline
100	&	0.83	&	0.83	&	5.9	&	5.2	&	8	&	2.6	\\
105	&	0.84	&	0.84	&	3.0	&	4.6	&	7	&	1.4	\\
110	&	0.93	&	0.88	&	1.7	&	5.5	&	3	&	0.7	\\
115	&	0.93	&	0.88	&	0.9	&	5.2	&	1	&	0.4	\\
120	&	1.02	&	0.87	&	0.6	&	4.4	&	2	&	0.3	\\
125	&	1.02	&	0.87	&	0.4	&	4.0	&	2	&	0.2	\\
130	&	1.06	&	0.89	&	0.3	&	2.8	&	4	&	0.2	\\
135	&	1.05	&	0.88	&	0.2	&	2.2	&	5	&	0.1	\\
140	&	1.12	&	0.90	&	0.1	&	2.2	&	3	&	0.06	\\
145	&	1.14	&	0.86	&	0.04	&	2.0	&	2	&	0.03	\\
150	&	1.14	&	0.95	&	0.02	&	2.0	&	1	&	0.02	\\      \hline\hline 
   \end{tabular} }
   \label{tab:EventYieldsCCConv_High} 
   }\hspace{0.6em}
     \subtable[]{
\scalebox{1.0}{ 
      \begin{tabular}{ccccccc} 
      \hline\hline\\[-2ex]
\multicolumn{4}{l}{C$^{\prime}$P Channel, $p_T^{\gamma\gamma}>$~75~GeV} &   \multicolumn{3}{r}{CDF Run II, 6.7 fb$^{-1}$}    \\
\hline\\[-2.3ex]
      $m_{h_f}$ & \multicolumn{2}{c}{$\epsilon$A (\%)} & $h_{f}$ &  &  & \\
      (GeV/$c^{2}$) & VH & VBF & Signal & Background & Data & $S$/$\sqrt{B}$\\
      \hline
100	&	0.27	&	0.28	&	1.9	&	8.0	&	6	&	0.7	\\
105	&	0.31	&	0.30	&	1.1	&	7.6	&	5	&	0.4	\\
110	&	0.35	&	0.31	&	0.6	&	6.0	&	7	&	0.2	\\
115	&	0.35	&	0.31	&	0.3	&	5.9	&	5	&	0.1	\\
120	&	0.40	&	0.33	&	0.2	&	6.0	&	4	&	0.09	\\
125	&	0.41	&	0.33	&	0.2	&	5.6	&	4	&	0.07	\\
130	&	0.42	&	0.36	&	0.1	&	5.6	&	2	&	0.05	\\
135	&	0.46	&	0.38	&	0.08	&	5.0	&	2	&	0.03	\\
140	&	0.47	&	0.39	&	0.03	&	4.3	&	4	&	0.02	\\
145	&	0.48	&	0.38	&	0.02	&	3.5	&	5	&	0.01	\\
150	&	0.51	&	0.40	&	0.01	&	2.7	&	7	&	0.01	\\      \hline \hline
   \end{tabular} }
   \label{tab:EventYieldsCPConv_High} 
   }}
   \end{centering}
   \end{table} 

%%%%%%%%%%%%%%%%%%%%%%%%%%%%%%%%%%%%%%%%%%%%%%%%%
%                                         Medium Pt Bin
%%%%%%%%%%%%%%%%%%%%%%%%%%%%%%%%%%%%%%%%%%%%%%%%%  
  \newpage
  \subsection{(B) 35~$<p_T^{\gamma\gamma}<$~75~GeV/$c$ Region}

          \begin{table}[htb] 
   \caption{The fermiophobic Higgs boson search is divided into four independent categories (CC, CP, C$^{\prime}$C, and C$^{\prime}$P) 
   as in the SM Higgs boson search, and it is additionally divided into three regions of 
   $p_T^{\gamma\gamma}$.  The 35~$<p_T^{\gamma\gamma}<$~75~GeV/$c$ region is shown here. 
   For each $h_f$ mass hypotheses tested in this analysis, the efficiency 
   multiplied by signal acceptance ($\epsilon$A) is shown as a percentage of the total number 
   of \Hgg\ decays for each production mechanism (VH and VBF).
   These values, along with the cross sections and branching ratios provided 
   in Table~\ref*{tab:sigma}, are used to obtain the predicted number of SM Higgs boson
   signal events. The CC and C$^{\prime}$C (CP and C$^{\prime}$P) channels use 
   data corresponding to an integrated luminosity of 7.0~fb$^{-1}$ (6.7~fb$^{-1}$). 
   The number of background and data events are also given for each mass. 
   The final column in each subtable is the number of signal events divided by the
   square root of the number of background events ($S$/$\sqrt{B}$). 
   The event yields for each mass point are 
   obtained from a 12~GeV/$c^{2}$ signal region centered on the Higgs boson mass
   hypothesis, allowing a 2~GeV/$c^{2}$ overlap between signal regions.}  
   \label{Tab:EventYields_Med} 
      \begin{centering}
\makebox[\textwidth]{
      \subtable[]{
     \scalebox{1.0}{ 
      \begin{tabular}{ccccccc}
      \hline\hline\\[-2ex] 
\multicolumn{7}{l}{CC Channel, 35 $<p_T^{\gamma\gamma}<$~75~GeV/$c$ \quad CDF Run II, 7.0 fb$^{-1}$} \\
\hline\\[-2.3ex]
      $m_{h_f}$ & \multicolumn{2}{c}{$\epsilon$A (\%)} & $h_{f}$ &  &  & \\
      (GeV/$c^{2}$) & VH & VBF & Signal & Background & Data & $S$/$\sqrt{B}$\\
      \hline
100	&	4.00	&	4.61	&	29.4	&	97.0	&	101	&	3.0	\\
105	&	3.87	&	4.55	&	14.4	&	83.2	&	78	&	1.6	\\
110	&	3.87	&	4.62	&	7.4	&	64.7	&	69	&	0.9	\\
115	&	3.83	&	4.63	&	3.8	&	57.7	&	51	&	0.5	\\
120	&	3.78	&	4.60	&	2.4	&	43.9	&	47	&	0.4	\\
125	&	3.66	&	4.57	&	1.6	&	37.4	&	41	&	0.3	\\
130	&	3.67	&	4.63	&	1.3	&	30.5	&	31	&	0.2	\\
135	&	3.56	&	4.58	&	0.7	&	27.6	&	24	&	0.1	\\
140	&	3.52	&	4.60	&	0.3	&	22.7	&	18	&	0.06	\\
145	&	3.43	&	4.51	&	0.1	&	18.3	&	18	&	0.03	\\
150	&	3.41	&	4.57	&	0.08	&	15.1	&	18	&	0.02	\\      \hline \hline
   \end{tabular} }
   \label{tab:EventYieldsCC_Med} 
  }\hspace{0.6em}
  \subtable[]{
\scalebox{1.0}{ 
      \begin{tabular}{ccccccc} 
      \hline\hline\\[-2ex] 
\multicolumn{7}{l}{CP Channel, 35 $<p_T^{\gamma\gamma}<$~75~GeV/$c$ \quad CDF Run II, 6.7 fb$^{-1}$} \\
\hline\\[-2.3ex]
      $m_{h_f}$ & \multicolumn{2}{c}{$\epsilon$A (\%)} & $h_{f}$ &  &  & \\
      (GeV/$c^{2}$) & VH & VBF & Signal & Background & Data & $S$/$\sqrt{B}$\\
      \hline
100	&	4.80	&	4.91	&	33.0	&	546	&	554	&	1.4	\\
105	&	4.88	&	5.01	&	16.9	&	501	&	485	&	0.8	\\
110	&	4.84	&	4.97	&	8.5	&	439	&	413	&	0.4	\\
115	&	4.77	&	5.03	&	4.4	&	398	&	362	&	0.2	\\
120	&	4.75	&	5.01	&	2.8	&	344	&	316	&	0.2	\\
125	&	4.63	&	5.08	&	1.9	&	312	&	275	&	0.1	\\
130	&	4.55	&	4.97	&	1.4	&	263	&	260	&	0.09	\\
135	&	4.46	&	4.99	&	0.8	&	232	&	253	&	0.05	\\
140	&	4.33	&	4.96	&	0.3	&	203	&	193	&	0.02	\\
145	&	4.30	&	4.84	&	0.2	&	182	&	180	&	0.01	\\
150	&	4.17	&	4.74	&	0.1	&	154	&	158	&	0.01	\\      \hline \hline
   \end{tabular}} 
   \label{tab:EventYieldsCP_Med} 
   }}\vspace{5 mm}
   \makebox[\textwidth]{
  \subtable[]{
\scalebox{1.0}{ 
      \begin{tabular}{ccccccc} 
      \hline\hline\\[-2ex] 
\multicolumn{7}{l}{C$^{\prime}$C Channel, 35 $<p_T^{\gamma\gamma}<$~75~GeV/$c$ \quad CDF Run II, 7.0 fb$^{-1}$} \\
\hline\\[-2.3ex]
      $m_{h_f}$ & \multicolumn{2}{c}{$\epsilon$A (\%)} & $h_{f}$ &  &  & \\
      (GeV/$c^{2}$) & VH & VBF & Signal & Background & Data & $S$/$\sqrt{B}$\\
      \hline
100	&	0.87	&	1.00	&	6.4	&	21.3	&	21	&	1.4	\\
105	&	0.86	&	1.03	&	3.2	&	18.9	&	19	&	0.7	\\
110	&	0.86	&	1.05	&	1.6	&	16.2	&	14	&	0.4	\\
115	&	0.85	&	1.05	&	0.8	&	13.9	&	13	&	0.2	\\
120	&	0.87	&	1.04	&	0.6	&	10.2	&	16	&	0.2	\\
125	&	0.82	&	1.03	&	0.4	&	9.4	&	12	&	0.1	\\
130	&	0.83	&	1.09	&	0.3	&	8.8	&	7	&	0.1	\\
135	&	0.79	&	1.02	&	0.2	&	7.9	&	4	&	0.06	\\
140	&	0.80	&	1.04	&	0.07	&	6.4	&	5	&	0.03	\\
145	&	0.74	&	1.04	&	0.03	&	5.5	&	5	&	0.01	\\
150	&	0.76	&	1.04	&	0.02	&	4.0	&	6	&	0.01	\\      \hline \hline
   \end{tabular} }
   \label{tab:EventYieldsCCConv_Med} 
   }\hspace{0.6em}
     \subtable[]{
\scalebox{1.0}{ 
      \begin{tabular}{ccccccc} 
      \hline\hline\\[-2ex] 
\multicolumn{7}{l}{C$^{\prime}$P Channel, 35 $<p_T^{\gamma\gamma}<$~75~GeV/$c$ \quad CDF Run II, 6.7 fb$^{-1}$} \\
\hline\\[-2.3ex]
      $m_{h_f}$ & \multicolumn{2}{c}{$\epsilon$A (\%)} & $h_{f}$ &  &  & \\
      (GeV/$c^{2}$) & VH & VBF & Signal & Background & Data & $S$/$\sqrt{B}$\\
      \hline
100	&	0.46	&	0.47	&	3.16	&	77.7	&	73	&	0.36	\\
105	&	0.50	&	0.53	&	1.75	&	65.0	&	57	&	0.22	\\
110	&	0.48	&	0.50	&	0.85	&	51.7	&	50	&	0.12	\\
115	&	0.48	&	0.53	&	0.45	&	42.7	&	50	&	0.07	\\
120	&	0.48	&	0.52	&	0.29	&	32.3	&	50	&	0.05	\\
125	&	0.50	&	0.51	&	0.20	&	32.9	&	42	&	0.04	\\
130	&	0.49	&	0.55	&	0.16	&	31.4	&	25	&	0.03	\\
135	&	0.46	&	0.52	&	0.08	&	28.5	&	22	&	0.02	\\
140	&	0.45	&	0.52	&	0.04	&	25.6	&	13	&	0.01	\\
145	&	0.45	&	0.54	&	0.02	&	20.6	&	16	&	0.004	\\
150	&	0.46	&	0.52	&	0.01	&	16.1	&	20	&	0.002	\\      \hline \hline
   \end{tabular} }
   \label{tab:EventYieldsCPConv_Med} 
   }}
   \end{centering}
   \end{table} 
  
%%%%%%%%%%%%%%%%%%%%%%%%%%%%%%%%%%%%%%%%%%%%%%%%%
%                                         Low Pt Bin
%%%%%%%%%%%%%%%%%%%%%%%%%%%%%%%%%%%%%%%%%%%%%%%%%  
    \newpage
  \subsection{(C) $p_T^{\gamma\gamma}<$ 35 GeV/$c$ Region}
   \begin{table}[htb] 
 \caption{The fermiophobic Higgs boson search is divided into four independent categories (CC, CP, C$^{\prime}$C, and C$^{\prime}$P) 
   as in the SM Higgs boson search, and it is additionally divided into three regions of 
   $p_T^{\gamma\gamma}$.  The $p_T^{\gamma\gamma}<$ 35 GeV/$c$ region is shown here.  
   For each $h_f$ mass hypotheses tested in this analysis, the efficiency 
   multiplied by signal acceptance ($\epsilon$A) is shown as a percentage of the total number 
   of \Hgg\ decays for each production mechanism (VH and VBF).
   These values, along with the cross sections and branching ratios provided 
   in Table~\ref*{tab:sigma}, are used to obtain the predicted number of SM Higgs boson
   signal events. The CC and C$^{\prime}$C (CP and C$^{\prime}$P) channels use 
   data corresponding to an integrated luminosity of 7.0~fb$^{-1}$ (6.7~fb$^{-1}$). 
   The number of background and data events are also given for each mass. 
   The final column in each subtable is the number of signal events divided by the
   square root of the number of background events ($S$/$\sqrt{B}$). 
   The event yields for each mass point are 
   obtained from a 12~GeV/$c^{2}$ signal region centered on the Higgs boson mass
   hypothesis, allowing a 2~GeV/$c^{2}$ overlap between signal regions.}  
   \label{Tab:EventYields_Low} 
      \begin{centering}
\makebox[\textwidth]{
      \subtable[]{
     \scalebox{1.0}{ 
      \begin{tabular}{ccccccc} 
      \hline\hline\\[-2ex] 
\multicolumn{4}{l}{CC Channel, $p_T^{\gamma\gamma}<$ 35 GeV/$c$} & \multicolumn{3}{r}{CDF Run II, 7.0 fb$^{-1}$}    \\
\hline\\[-2.3ex]
      $m_{h_f}$ & \multicolumn{2}{c}{$\epsilon$A (\%)} & $h_{f}$ &  &  & \\
      (GeV/$c^{2}$) & VH & VBF & Signal & Background & Data & $S$/$\sqrt{B}$\\
      \hline
100	&	2.30	&	2.66	&	16.9	&	490	&	482	&	0.76	\\
105	&	2.15	&	2.67	&	8.1	&	413	&	397	&	0.40	\\
110	&	2.09	&	2.64	&	4.0	&	328	&	308	&	0.22	\\
115	&	2.01	&	2.62	&	2.0	&	269	&	289	&	0.12	\\
120	&	1.92	&	2.60	&	1.3	&	214	&	240	&	0.09	\\
125	&	1.84	&	2.51	&	0.8	&	189	&	215	&	0.06	\\
130	&	1.77	&	2.55	&	0.6	&	156	&	160	&	0.05	\\
135	&	1.73	&	2.53	&	0.4	&	138	&	147	&	0.03	\\
140	&	1.64	&	2.54	&	0.2	&	110	&	125	&	0.01	\\
145	&	1.58	&	2.50	&	0.07	&	100	&	103	&	0.007	\\
150	&	1.52	&	2.50	&	0.04	&	88	&	74	&	0.004	\\      \hline \hline
   \end{tabular} }
   \label{tab:EventYieldsCC_Low} 
  }\hspace{0.3em}
  \subtable[]{
\scalebox{1.0}{ 
      \begin{tabular}{ccccccc} 
      \hline\hline\\[-2ex] 
\multicolumn{4}{l}{CP Channel, $p_T^{\gamma\gamma}<$ 35 GeV/$c$} & \multicolumn{3}{r}{CDF Run II, 6.7 fb$^{-1}$}    \\
\hline\\[-2.3ex]
      $m_{h_f}$ & \multicolumn{2}{c}{$\epsilon$A (\%)} & $h_{f}$ &  &  & \\
      (GeV/$c^{2}$) & VH & VBF & Signal & Background & Data & $S$/$\sqrt{B}$\\
      \hline
100	&	2.92	&	3.08	&	20.2	&	3646	&	3614	&	0.33	\\
105	&	2.78	&	3.12	&	9.8	&	3014	&	3055	&	0.18	\\
110	&	2.73	&	3.07	&	4.9	&	2423	&	2375	&	0.10	\\
115	&	2.62	&	3.07	&	2.5	&	2100	&	2052	&	0.05	\\
120	&	2.48	&	3.02	&	1.5	&	1683	&	1709	&	0.04	\\
125	&	2.38	&	2.98	&	1.0	&	1463	&	1556	&	0.03	\\
130	&	2.28	&	2.94	&	0.8	&	1215	&	1200	&	0.02	\\
135	&	2.23	&	2.88	&	0.4	&	1071	&	1055	&	0.01	\\
140	&	2.06	&	2.82	&	0.2	&	886	&	849	&	0.006	\\
145	&	2.00	&	2.80	&	0.08	&	782	&	742	&	0.003	\\
150	&	1.92	&	2.76	&	0.04	&	643	&	657	&	0.002	\\      \hline \hline
   \end{tabular}} 
   \label{tab:EventYieldsCP_Low} 
   }}\vspace{5 mm}
   \makebox[\textwidth]{
  \subtable[]{
\scalebox{1.0}{ 
      \begin{tabular}{ccccccc} 
      \hline\hline\\[-2ex] 
\multicolumn{4}{l}{C$^{\prime}$C Channel, $p_T^{\gamma\gamma}<$ 35 GeV/$c$} &   \multicolumn{3}{r}{CDF Run II, 7.0 fb$^{-1}$}    \\
\hline\\[-2.3ex]
      $m_{h_f}$ & \multicolumn{2}{c}{$\epsilon$A (\%)} & $h_{f}$ &  &  & \\
      (GeV/$c^{2}$) & VH & VBF & Signal & Background & Data & $S$/$\sqrt{B}$\\
      \hline
100	&	0.51	&	0.62	&	3.8	&	130	&	129	&	0.33	\\
105	&	0.51	&	0.61	&	1.9	&	99.6	&	115	&	0.19	\\
110	&	0.48	&	0.61	&	0.9	&	82.5	&	75	&	0.10	\\
115	&	0.43	&	0.57	&	0.4	&	66.6	&	66	&	0.05	\\
120	&	0.43	&	0.61	&	0.3	&	52.4	&	49	&	0.04	\\
125	&	0.42	&	0.57	&	0.2	&	45.8	&	41	&	0.03	\\
130	&	0.42	&	0.62	&	0.2	&	36.9	&	33	&	0.03	\\
135	&	0.38	&	0.59	&	0.08	&	32.2	&	30	&	0.01	\\
140	&	0.39	&	0.58	&	0.04	&	24.6	&	32	&	0.007	\\
145	&	0.37	&	0.57	&	0.02	&	23.0	&	31	&	0.003	\\
150	&	0.36	&	0.58	&	0.01	&	22.4	&	16	&	0.002	\\      \hline \hline
   \end{tabular} }
   \label{tab:EventYieldsCCConv_Low} 
   }\hspace{0.3em}
     \subtable[]{
\scalebox{1.0}{ 
      \begin{tabular}{ccccccc} 
      \hline\hline\\[-2ex] 
\multicolumn{4}{l}{C$^{\prime}$P Channel, $p_T^{\gamma\gamma}<$ 35 GeV/$c$} &   \multicolumn{3}{r}{CDF Run II, 6.7 fb$^{-1}$}    \\
\hline\\[-2.3ex]
      $m_{h_f}$ & \multicolumn{2}{c}{$\epsilon$A (\%)} & $h_{f}$ &  &  & \\
      (GeV/$c^{2}$) & VH & VBF & Signal & Background & Data & $S$/$\sqrt{B}$\\
      \hline
100	&	0.29	&	0.34	&	2.0	&	557	&	573	&	0.09	\\
105	&	0.31	&	0.35	&	1.1	&	395	&	381	&	0.06	\\
110	&	0.28	&	0.34	&	0.5	&	274	&	299	&	0.03	\\
115	&	0.28	&	0.33	&	0.3	&	234	&	263	&	0.02	\\
120	&	0.27	&	0.33	&	0.2	&	188	&	213	&	0.01	\\
125	&	0.27	&	0.33	&	0.1	&	167	&	188	&	0.009	\\
130	&	0.26	&	0.33	&	0.09	&	139	&	153	&	0.007	\\
135	&	0.24	&	0.32	&	0.05	&	127	&	116	&	0.004	\\
140	&	0.23	&	0.33	&	0.02	&	109	&	89	&	0.002	\\
145	&	0.23	&	0.33	&	0.009	&	95	&	83	&	0.001	\\
150	&	0.22	&	0.31	&	0.005	&	80	&	72	&	0.001	\\      \hline \hline
   \end{tabular} }
   \label{tab:EventYieldsCPConv_Low} 
   }}
   \end{centering}
   \end{table} 

\end{document}